\documentclass{kluwer}
\begin{document}
\begin{article}
\begin{opening}
\title{
Nature of the compact X-ray source in supernova remnant RCW103 and related problems}
\author{Sergei B. \surname{Popov}}
\runningauthor{Sergei Popov}
\runningtitle{Compact X-ray source in RCW103}
\institute{Sternberg Astronomical Institute, Moscow, Russia}
\date{}

\begin{abstract}
I discuss the nature of the compact X-ray source in the center of
the supernova remnant RCW 103. Several models, based on the accretion
onto a compact object are analyzed.
I show that it is more likely that the central X-ray source is an
accreting neutron star than an accreting black hole.
I also argue that models of a disrupted
binary system consisting of  an old accreting neutron star
and a new one observed as a 69-ms  pulsar are most favored. 
\end{abstract}

\end{opening}

\section{Introduction}

Among all astrophysical objects neutron stars (NSs) and black holes (BHs)
attract most attention of physicists.
Now we know more than 1000 NSs as 
radio pulsars and more than 100 NSs emitting X- or/and $\gamma$-rays.
The Galactic population of these objects is much larger: about $10^8$ -- $10^9$.

It is generally accepted that NSs
and BHs are the products
of  supernova (SN) explosions.
 In most cases a supernova remnant (SNR)
appears after a {\it formidable} explosion of a massive star 
(with $M > 10-35 M_{\odot}$).
Although sometimes a young NS is observed inside a SNR 
as a radio pulsar (e.g., Crab, Vela, etc.) or as a X-ray source, 
in most cases no compact object is found inside a SNR, or an
accidental coincidence
of the radio pulsar and the SNR is very likely (e.g.,
\cite{kaspi98}; \cite{frail}).

It is possible, that about 50\% of NSs are born with low
magnetic fields, so they never appear as radio pulsars and spend
most of their lives on the Ejector and Propeller stages. 
These NSs with low magnetic fields 
can not spin-down significantly even 
during the Hubble time.
For high-velocity NSs the characteristic Ejector period is higher,
and NSs spend most of their lives as Ejectors.

Recently, \citeauthor{gotthelf} described a compact X-ray source 
in the center of SNR RCW 103 with the 
X-ray luminosity $L_x\sim 10^{34}\, {\rm erg}\,{\rm s}^{-1}$ 
(for the distance $3.3\, {\rm kpc}$)
and the black-body temperature  about $0.6\, {\rm keV}$.
The source flux has varied since previous observations \cite{petre}.
The nature of the central compact source is unclear. 
No radio or optical compact counterpart was observed. 
Also a 69-ms X-ray and radio pulsar with a characteristic age
about 8 kyr was discovered 7' from the
center of the remnant \cite{kaspi98},
but the reality of the association
of the pulsar with the SNR is unclear \cite{dickel}.

Here I discuss possible models of that compact central source
and its possible connections with the 69-ms pulsar (see some preliminary
results in \cite{popov98}).

\section{What is inside the RCW 103?}

\citeauthor{gotthelf} discussed why the source cannot be
a cooling NS, a plerion, or a binary with a normal companion.
The reader is referred to their paper for the details.
In the present analysis I assume that the X-ray luminosity of the 
source is  produced due to accretion of the surrounding material 
onto a compact object (a NS or a  BH). 
I analyze thus only models with compact objects, isolated or with a compact
companion (most probably the binary system was destroyed after the second
explosion, when the 69-ms pulsar was formed). 

 The main challenge for the models of accretion of the surrounding material
onto isolated compact object is to answer the question of
where a NS or a BH finds enough matter to accrete. 
I don't discuss it here, assuming that
the material is available in the surrounding medium (see,
for example, \cite{page}).

\subsection{Accreting isolated young black hole or accreting
old black hole in pair with a young compact object}

An isolated BH accreting the interstellar medium can be, 
in principle, observed
by X-ray satellites such as $ROSAT$, $ASCA$ etc. \cite{heckler}.
To achieve high X-ray luminosity, 
a compact object must move with a low velocity relative the ISM:

\begin{equation}
 \dot M=2 \pi \left(
    \frac{(GM)^2 \rho}{(V_s^2+V^2)^{3/2}}\right),
\end{equation}
where $V_s$ is the speed of sound, $V$ is the velocity of the
compact object with respect to the ambient medium, $M$-- the mass of the 
accreting star and $\rho$ is the density of the accreting material.
One can introduce the effective velocity, $V_{eff}$, and rewrite eq. (1)
as follows:
$$
\dot M=2 \pi ((GM)^2 \rho)/(V_{eff}^3).
$$

During the SN explosions a compact object can obtain an additional 
kick velocity. At the present time the distribution of the kick velocity is 
not known well enough (e.g., \cite{lpp}).
Although observations of radio pulsars favor high kick velocities 
about $300-\, 500 \, {\rm km}\, {\rm s}^{-1}$
\cite{lyne}.
We mark here, that if the 69-ms pulsar is a new born NS,
and the central source is the older object, it is not surprising,
that the 69-ms pulsar is farther from the center of the SNR. Because
the new born NS received a high kick velocity
and the old one only saved its orbital velocity, because the system survived
in the first explosion. 
X-ray radiation of the new born NS
of course doesn't have the accretion nature.

To explain the observed X-ray luminosity of the compact object in
the center of RCW 103 
 the accretion rate, $\dot M$, should be about $10^{14}\, {\rm g}\, {\rm
s}^{-1}$.

One can then estimate the size of the emmiting region, 
using observed luminosity and temperature: $
 L=4\pi \cdot R_{emm}^2 \sigma T^4 $.
For observed values of $ L_x$ and $ T$ 
this equation gives $R_{emm} \sim 1 \,{\rm km}$.
For BHs such a low value of $R_{emm}$ is very unlikely because
the gravitational radius is about $R_G\sim 3\,{\rm km}\, (M/M_{\odot})$,
and most of the present BH-candidates have masses about $7-10 M_{\odot}$.
This is probably the main argument against isolated accreting
BH as a model for the RCW 103. 
Also the efficiency of spherically symmetric accretion onto a
BH is very low resulting in a significantly higher density required
to achieve the same luminosity.
The same arguments can be used against models with a binary system
(probably disrupted): BH+NS
(NS was born in the recent SN explosion -- a 69-ms pulsar).

\subsection{Young isolated accreting neutron star}

 In the past few years isolated accreting NSs have become a subject 
of great interest especially due to the
observations with the $ROSAT$ satellite.

 There are four main possible stages for a NS in a low-density plasma:
$1).$ Ejector (a radio pulsar is an example of Ejector); $2).$ Propeller; 
 $3).$ Accretor; and $4).$ Georotator
(\cite{lp}; \cite{konenkov}).
 The stage is determined by the accretion rate, $\dot M$, the magnetic field
of the NS, $B$, and by the spin period of the NS, $p$.

 If the NS is on the Accretor stage, then its period is longer than the
so-called Accretor period, $P_A$:

\begin{equation}
 P_A=2^{5/14}\pi \, (GM)^{-5/7} (\mu ^2/\dot M)^{3/7}\, {\rm s},
\end{equation} 
where $\mu = B\cdot R_{NS}^3$ is magnetic moment of the NS.

For the RCW 103 I use the following values: $\dot M= 10^{14}\, {\rm g}\,
{\rm s}^{-1}$,
$M=1.4\, M_{\odot}$, $R_{NS}=10^6\, {\rm cm}$ which give:

\begin{equation}
B\sim 10^{10}\cdot p^{7/6}\, {\rm G}.
\end{equation}

If material is accreted from the turbulent interstellar medium, a new
equilibrium period can occur \cite{konenkov}:

\begin{equation}
	P_{eq}\sim 30\, B_{12}^{2/3}I_{45}^{1/3}\dot M_{14}^{-2/3}
R_{{NS}_6}^2 V_{{eff}_6}^{7/3}V_{{t}_6}^{-2/3}M_{1.4}^{-4/3}\, {\rm s},
\end{equation}
where $V_t$ is the turbulent velocity (all velocities are in units 
of $10\, {\rm km} \, {\rm s}^{-1}$); 
$M_{1.4}$ is the  mass of the NS in units of $1.4\, M_{\odot}$,
$B_{12}$ is the magnetic field of the NS in unites
$10^{12}\, {\rm G}$ and $R_{NS}$ is the radius of the NS in units of $10^6\, {\rm
cm}$.

We then obtain:

\begin{equation}
  B\sim 8\cdot 10^{9}\cdot p^{3/2}\, {\rm G}.
\end{equation}

It is obvious that to explain the luminosity 
of the RCW 103 by an isolated accreting NS,
one must assume that the NS was born with extremely low magnetic 
field  or with
unusually long spin period. The age of the SNR RCW 103 is about 1000 years
\cite{gotthelf}, which
means that the magnetic field could not decay significantly \cite{konenkov}. 
The flux of the source is not constant \cite{petre}, so the idea
of cooling NS can be rejected. Thus, the model with 
isolated young accreting NS is not a likely explanation for the data.

\subsection{Accreting old neutron star in pair 
with a young neutron star (or in the disrupted system)}

 Binary compact objects are quite natural products of binary evolution
\cite{lpp}. One can, therefore, discuss these scenarios
as a viable alternative.

 In the previous subsection I showed that accretion onto a young isolated
NS requires unusual initial parameters. However, 
there is a chance that we observe a binary system
(or a disrupted binary), where one component is an 
old NS and the other component was formed in a recent SN explosion
and appears (for disrupted system) as a 69-ms pulsar.

 In that case, the parameters 
determined by eqs.(3), (5) are not unusual:
old NS can have low magnetic fields and long periods 
\cite{lp}.
Due to the fact
that  \citeauthor{gotthelf} did not find any periodic change of the 
luminosity, one can argue that the field is too low to produce
the observable modulation (the accreting material is not
channeled to the polar caps: $B <10^6\, {\rm G}$) or that the period is very long
($p>10^4\, {\rm s}$), which is possible  for old NSs with ``normal'' 
magnetic fields \cite{lp}:
$
P\approx 500\, {\rm s}.
$
The last opportunity is, probably, better, as the emmiting area
is not large $\approx 1 \, {\rm km}^2$.

 The evolutionary scenario for such a system is clear enough \cite{lpp}. 
One can easily calculate it using the ``Scenario Machine''
WWW-facility \cite{nazin}. For example,
two stars with masses $15 \, M_{\odot}$ and $14 \, M_{\odot}$ on the
main sequence with the initial separation $200 \, R_{\odot}$, $R_{\odot}$ --
the solar radius, after 14 Myr  (with two SN explosions
with low kick velocities: about $60 \, {\rm km}\, {\rm s}^{-1}$) 
end their evolution as a binary system NS+NS.
The second NS is $\sim 1$ Myr younger. During $\sim 1$ 
Myr the magnetic field can decrease
up to 1/100 of the initial value with a significant spin-down 
\cite{konenkov}.
The binary NS+NS is relatively wide: $ 20\, R_{\odot}$ with an 
orbital period $5.8^{\rm d}$, so the orbital velocity is not high (the orbital
velocity of the accreting NS should be added to $V_{eff}$).

The 69-ms X-ray pulsing source and it's radio pulsar counterpart that
were discovered near RCW 103 \cite{kaspi98} 
can be a new-born radio pulsar.
So, it means that the binary system was disrupted after the second explosion.
It means that in the first explosion the kick velocity was small
(about $50\, {\rm km}\, {\rm s}^{-1}$ 
in the opposite case the system could be disrupted after the
first explosion and the older NS could leave the SNR before
the second explosion, but if the orbit was significantly eccentric,
the kick velocity in the first explosion could be high too) 
and in the second explosion it was as high 
as $750-800 \, {\rm km}\, {\rm s}^{-1}$ 
for the same initial parameters as in the previous example.

Of course other variants of the initial parameters are possible,
and I showed this one just as a simple example.

\section{Conclusions}

To conclude, I argued that the most likely model for the 
central compact X-ray source of RCW 103 is
that of an accreting old NS in a disrupted binary system with a young
compact object (the 69-ms pulsar) born in the recent 
SN explosion that produced the observed supernova remnant
(some ideas about a disrupted binary 
in RCW 103 were also discussed in the article
\cite{torii}).
Such systems are rare, but natural products
of the binary evolution. Scenarios with a single compact objects or with
accreting BH are less probable.

\acknowledgements
I thank M.E. Prokhorov and V.M.Lipunov for helpful discussions,
E.V. Gotthelf and K. Torii for the information about RCW 103 and
A.V. Krav\-tsov for his comments on the text. 
The work was supported by 
the INTAS (96-0315) and NTP "Astronomy" (1.4.4.1) grants.

\end{article}
\end{document}